\documentstyle[preprint,aps]{revtex}
\tighten
\draft
\widetext
\input epsf
\preprint{HUTP-99/A012, NUB 3194}
\bigskip
\bigskip
\begin{document}
\title{Non-perturbative Orientifolds}
\medskip

\author{Zurab Kakushadze\footnote{E-mail: 
zurab@string.harvard.edu}}

\bigskip
\address{Jefferson Laboratory of Physics, Harvard University,
Cambridge,  MA 02138\\
and\\
Department of Physics, Northeastern University, Boston, MA 02115}

\date{April 2, 1999}
\bigskip
\medskip
\maketitle

\begin{abstract}
{}We consider non-perturbative six and four dimensional ${\cal N}=1$ space-time
supersymmetric orientifolds. Some states in such compactifications arise in
``twisted'' open string sectors which lack world-sheet description in terms
of D-branes. Using Type I-heterotic duality we are able to obtain the
massless spectra for some of such orientifolds. The four dimensional
compactification we discuss in this context is an example of a chiral 
${\cal N}=1$ supersymmetric 
string vacuum which is non-perturbative in both orientifold
and heterotic pictures. In particular, it contains both D9- and D5-branes
plus non-perturbative ``twisted'' open string sector states as well.

\end{abstract}

\section{Introduction}

{}In recent years substantial progress has been made in understanding
orientifold compactifications. Various six dimensional orientifold vacua
were constructed, for instance, in \cite{PS,GP,GJ,bij}. Generalizations of
these constructions to four dimensional orientifold vacua have also been
discussed in detail in \cite{BL,Sagnotti,KS,Zw,ibanez,KST}. In many cases
the world-sheet approach to orientifolds is adequate and gives rise to
consistent anomaly free vacua in six and four dimensions. However, it was
pointed out in \cite{KST} that there
are cases where the perturbative orientifold description misses certain
non-perturbative sectors which are nonetheless present and must be taken
into account. In certain cases this inadequacy results in obvious
inconsistencies such as lack of tadpole and anomaly cancellation. Examples
of such cases were discussed in \cite{Zw,ibanez,KST}. In other cases,
however, the issue is more subtle as the non-perturbative states arise in
anomaly free combinations, so that they are easier to miss. 

{}In certain four dimensional ${\cal N}=1$ supersymmetric cases 
Type I-heterotic duality enables one to argue along the lines of 
\cite{ZK,KS,KST} that the non-perturbative states become heavy 
and decouple once the 
corresponding orbifold singularities are appropriately blown
up. In particular, these are the ${\bf Z}_3$ \cite{Sagnotti,ZK}, 
${\bf Z}_7$, ${\bf Z}_3\otimes {\bf Z}_3$ and ${\bf Z}_6$ \cite{KS},
${\bf Z}_2\otimes {\bf Z}_2\otimes {\bf Z}_3$ \cite{223}, and
$\Delta(3\cdot3^2)$ (the latter is non-Abelian) \cite{class}
orbifold cases. As was argued in \cite{KST}, the same, however, does not 
seem to be the case for 
other tadpole free orientifolds such as the ${\bf Z}_2\otimes{\bf Z}_6$,     
${\bf Z}_3\otimes{\bf Z}_6$ and ${\bf Z}_6\otimes{\bf Z}_6$ \cite{Zw}, and 
${\bf Z}_{12}$ \cite{ibanez} 
orbifold cases. The purpose of this paper is to understand
non-perturbative states in such orientifold compactifications. In
particular, in some cases we are able to obtain the massless states arising
in such sectors. 

{}The origin of non-perturbative states in orientifold compactifications
can already be understood in six dimensions. Thus, in the K3 orbifold examples 
of \cite{GJ} the orientifold projection is not $\Omega$, which we will use
to denote that in the smooth K3 case, but rather $\Omega J^\prime$, where
$J^\prime$ maps the $g$ twisted sector to its conjugate $g^{-1}$ twisted
sector (assuming $g^2\not=1$) \cite{pol}. Geometrically this can be viewed
as a permutation of two ${\bf P}^1$'s associated with each fixed point of
the orbifold \cite{KST}. (More precisely, these ${\bf P}^1$'s correspond to
the orbifold blow-ups.) This is different from the orientifold projection
in the smooth case where (after blowing up) the orientifold projection
does {\em not} permute the two ${\bf P}^1$'s. In the case of the 
$\Omega J^\prime$ projection the ``twisted'' open string sectors corresponding
to the orientifold elements $\Omega J^\prime g$ are absent
\cite{blum,KST}. However, if the orientifold projection is $\Omega$, then
the ``twisted'' open string sectors corresponding
to the orientifold elements $\Omega g$ are present. In fact, these states
are non-perturbative from the orientifold viewpoint and are required for
gravitational anomaly cancellation in six dimensions. As we will see, in
certain cases Type I-heterotic duality will allow us to understand such
sectors.

{}In four dimensional orientifolds with ${\cal N}=1$ supersymmetry there
are always sectors (except in the ${\bf Z}_2\otimes {\bf Z}_2$ model of
\cite{BL} which is completely perturbative from the orientifold viewpoint)
such that there is only one ${\bf P}^1$ per fixed point, so only the
$\Omega$ orientifold projection is allowed. This results in
non-perturbative ``twisted'' open string sectors, which, as we have
already mentioned, decouple in certain cases once the appropriate blow-ups 
are performed. In other cases we must include these states to obtain the
complete description of a given orientifold.      

{}Some of the non-perturbative orientifolds have perturbative heterotic
duals. However, non-perturbative orientifolds with, say, D5-branes are
non-perturbative from the heterotic viewpoint. In this paper we are
therefore exploring some vacua in the region ${\cal D}$ in Fig.1 (which has
been borrowed from \cite{class}). Thus, the ${\bf Z}_6^\prime$ orbifold
compactification we discuss in section III is an example of a four
dimensional {\em chiral} ${\cal N}=1$ supersymmetric string vacuum which is
non-perturbative from both orientifold and heterotic points of view.
    
{}In the next section we give examples of six dimensional non-perturbative
orientifolds. These are illustrative as the gravitational anomaly
cancellation requirement in six dimensions is rather constraining which 
makes it easier to discuss such vacua. Once we understand these six
dimensional examples, we move down to four dimensions in section III. 

\section{Six Dimensional Examples}

{}In this section we will give an example of a non-perturbative six
dimensional orientifold with ${\cal N}=1$ supersymmetry. This vacuum
corresponds to the $\Omega$ orientifold of Type IIB on the ${\bf Z}_6$
orbifold limit of K3 (that is, K3 in this case is $T^4/{\bf
Z}_6$). Alternatively, this model can be viewed as a Type I
compactification on the same K3 with D5-branes. In the dual heterotic
picture these D5-branes are mapped to small instantons.

{}Thus, consider Type IIB on ${\cal M}_2=T^4/{\bf Z}_6$ (with trivial NS-NS
$B$-field), where the
generator $g$ of ${\bf Z}_6(\approx{\bf Z}_3\otimes {\bf Z}_2)$ acts on 
the complex coordinates $z_1,z_2$ on 
$T^4$ as follows: 
\begin{equation}
 g z_1=\omega z_1~,~~~g z_2=\omega^{-1} z_2~,
\end{equation}
where $\omega\equiv\exp(2\pi i/6)$. This theory has ${\cal N}=2$
supersymmetry in six dimensions.

{}Next, we consider the $\Omega$ orientifold of this theory. Note that, as
we have already mentioned in the previous section, the $\Omega$ projection 
acts as in the smooth K3 case. This, in particular, implies that we must
first blow up the orbifold singularities before orientifolding.

{}Thus, after orientifolding the closed string sector contains only one
tensor multiplet (but no extra tensor multiplets), and 20 hypermultiplets
(plus the corresponding ${\cal N}=1$ supergravity multiplet). Here 2 of the
above hypermultiplets come from the untwisted sector, 2 more arise in the
${\bf Z}_6$ twisted sectors $g,g^{-1}$, $10(=2\times 5)$ hypermultiplets
come from the ${\bf Z}_3$ twisted sectors $g^2,g^{-2}$ (note that 5 linear
combinations of the original 9 points fixed under the ${\bf Z}_3$ action
are invariant under the ${\bf Z}_2$ twist), and, finally, 6 hypermultiplets
come from the ${\bf Z}_2$ twisted sector $g^3$ (note that 6 linear
combinations of the original 16 points fixed under the ${\bf Z}_2$ action
are invariant under the ${\bf Z}_3$ twist). 

{}As to the open string sector, there are 32 D9-branes plus 32
D5-branes. This is required by the untwisted tadpole cancellation
conditions. In the following we will consider the case where all 32
D5-branes are sitting on top of each other at one of the orientifold
5-planes which we choose to be that at the origin of the orbifold (that is,
at $z_1=z_2=0$). We can view the usual (``untwisted'') 99 and 55 open
string sectors as the ``$\Omega$-twisted'' sector of the orientifold. 
Similarly, the 59 sector can be viewed as ``$\Omega g^3$-twisted''
sector. These sectors are perturbative from the orientifold point of
view. That is, these sectors can be understood in the usual world-sheet
picture using the standard D-brane techniques. There are, however,
additional sectors in this model which do not possess such a simple
description. These are the ``$\Omega g^2$- and $\Omega g^{-2}$-twisted''
sectors, which can be viewed as the ${\bf Z}_3$ twisted 99 plus 55 sectors, as
well as the ``$\Omega g$- and $\Omega g^{-1}$-twisted'' sectors, which can be
viewed as the ${\bf Z}_3$ twisted 59 sector. These ${\bf Z}_3$ twisted open
string sectors are {\em non-perturbative} from the orientifold
viewpoint. Nonetheless, using Type I-heterotic duality we can (in this
particular example) understand these sectors and obtain the corresponding
massless states.   

{}The key observation here is the following. First, recall the ${\bf Z}_3$
model of \cite{GJ}. This model is obtained as follows. Consider Type IIB on 
${\cal M}_2=T^4/{\bf Z}_3$ (with trivial NS-NS $B$-field), where the
generator $\theta$ of ${\bf Z}_3$ (which is a subgroup of the above ${\bf
Z}_6$ orbifold group with the identification $\theta=g^2$) acts on 
the complex coordinates $z_1,z_2$ on $T^4$ as follows: 
\begin{equation}
 \theta z_1=\alpha z_1~,~~~\theta z_2=\alpha^{-1} z_2~,
\end{equation}
where $\alpha\equiv\omega^2=\exp(2\pi i/3)$. Next, consider the $\Omega
J^\prime$ orientifold of this theory, where $\Omega$ is the same as above
(that is, it acts as in the smooth K3 case), whereas $J^\prime$, which was
discussed in the previous section, maps the $\theta$ twisted sector to the   
$\theta^{-1}$ twisted sector \cite{pol}. After orientifolding the closed
string sector contains (along with the corresponding ${\cal N}=1$
supergravity multiplet) 10 tensor multiplets (one usual tensor multiplet
plus 9 extra tensor multiplets) and 11 hypermultiplets. In the open string
sector we have 32 D9-branes (but no D5-branes), and the action of the
orbifold group on the Chan-Paton charges is given by the corresponding
projective representation of ${\bf Z}_3$ in terms of the $16\times 16$
matrices\footnote{Throughout this paper we work with $16\times 16$
(rather than $32\times 32$) Chan-Paton matrices for we choose not to count the
orientifold images of the corresponding D-branes.}
$\{\gamma_1,\gamma_\theta,\gamma_{\theta^{-1}}\}$, where 
$\gamma_1={\bf I}_{16}$ is the
$16\times 16$ identity matrix, $\gamma_{\theta^{-1}}=\gamma^{-1}_\theta$, and
(up to equivalent representations) $\gamma_\theta$ is given by
\begin{equation}\label{Z_3}
 \gamma_\theta={\mbox{diag}}(\alpha{\bf I}_4,\alpha^{-1}{\bf I}_4,{\bf
 I}_8)~. 
\end{equation}      
Note that the above form of $\gamma_\theta$ is dictated by the
corresponding twisted tadpole cancellation conditions \cite{GJ}.
The massless spectrum of the open string sector of this model consists of
gauge vector 
supermultiplets of $U(8)\otimes SO(16)$ as well as the corresponding
charged hypermultiplets transforming in the irreps $({\bf 28},{\bf 1})$ and
$({\bf 8},{\bf 16})$ of the gauge group. Note that the above spectrum
satisfies the gravitational anomaly cancellation condition \cite{anomalies} 
\begin{equation}\label{anom}
 n_H-n_V=273-29 n_T~,
\end{equation}
where the number of tensor multiplets $n_T=10$, the number of
hypermultiplets $n_H=167$ (11 of which arise in the closed string sector,
while the other 156 come from the open string sector), and the number of
vector multiplets (all of which come from the open string sector) $n_V=184$.

{}The above model does not possess a perturbative heterotic dual as it
contains extra tensor multiplets. The reason for their appearance is that
the orientifold projection was chosen as $\Omega J^\prime$. On the other
hand, had we chosen the orientifold projection (after the appropriate
blow-ups) to be $\Omega$, we would have obtained a vacuum with no extra tensor
multiplets in the closed string sector. If, however, we choose the
Chan-Paton matrices to be the same as above (which in the 99 sector would
lead to the same massless spectrum), we would obtain an anomalous model
where the condition (\ref{anom}) is not satisfied. This signals that the
``naive'' orientifold (that is, world-sheet) approach in this case misses
some states whose existence is dictated by the anomaly cancellation
conditions. 

{}Fortunately, in the above case we can actually understand the origin of the
``missing'' states by using Type I-heterotic duality. The point is that
the $\Omega$ orientifold of Type IIB on the (appropriately blown up) 
orbifold $T^4/{\bf Z}_3$ is the same as Type I compactified on the same
space. The latter should be dual to a {\em perturbative} heterotic
compactification as we have no D5-branes in this case. In fact, we can
identify the corresponding heterotic vacuum as being the $T^4/{\bf Z}_3$
compactification with the gauge bundle defined by the orbifold action on
the ${\mbox{Spin}}(32)/{\bf Z}_2$ degrees of freedom given by the matrices
$\{{\bf W}_1,{\bf W}_\theta,{\bf
W}_{\theta^{-1}}\}\equiv\{\gamma_1,\gamma_\theta,\gamma_{\theta^{-1}}\}$.
Here the $16\times 16$ ${\bf W}$-matrices act on the
${\mbox{Spin}}(32)/{\bf Z}_2$ momentum states as the corresponding
rotations. (Alternatively, we can describe this action in terms of the
corresponding shifts of the ${\mbox{Spin}}(32)/{\bf Z}_2$ momentum lattice.)  
The resulting heterotic model is perturbatively well defined (that is, it
satisfies all the corresponding consistency requirements such as modular
invariance). The untwisted sector of this model contains the ${\cal N}=1$
supergravity multiplet, one tensor multiplet and 2 neutral hypermultiplets 
(these states are mapped to the untwisted closed string sector of the dual
Type I vacuum); it also contains the  
gauge vector supermultiplets of $U(8)\otimes SO(16)$ as well as 
the corresponding
charged hypermultiplets transforming in the irreps $({\bf 28},{\bf 1})$ and
$({\bf 8},{\bf 16})$ of the gauge group (these states are mapped to the
``untwisted'' open string sector of the dual Type I vacuum described above).
The twisted sector of the heterotic model contains 18 hypermultiplets
(neutral under the non-Abelian subgroups but fractionally\footnote{These
charges are fractional in the normalization (which is conventional in
discussing the orientifold models) where the fundamental of $SU(8)$ has
charge $+1$ under the corresponding $U(1)$ subgroup. In the following we
will omit the $U(1)$ charges which generically are fractional in the
``twisted'' open string sectors. In certain cases one can determine these 
charges using the Type I-heterotic duality (as on the heterotic side these
charges can be computed in the corresponding twisted sectors). Nonetheless,
we choose not to do this here as various subtleties involved in such
computations might have obscured the main point of this paper.} 
charged under the $U(1)$
subgroup) plus 9 additional hypermultiplets transforming in the irrep $({\bf
28},{\bf 1})$ of $U(8)\otimes SO(16)$ (these states are also fractionally
charged under the $U(1)$ subgroup). The above 18 hypermultiplets correspond
to the orbifold blow-up modes, and are mapped to the twisted closed string
sector hypermultiplets on the Type I side. Note that the 9 twisted 
hypermultiplets (on the heterotic side) charged under the $SU(8)$ subgroup
have no match in the orientifold picture. That is, these are
non-perturbative from the orientifold viewpoint. Nonetheless, they must
exist in the corresponding Type I model. In fact, they are precisely the
states required to satisfy the anomaly cancellation condition
(\ref{anom}). These states (which are charged under the 99 gauge group) can
be thought of as arising in ${\bf Z}_3$ twisted 99 sector. Type I-heterotic
duality is then a tool which allows us to understand such states (by doing a
perturbative heterotic calculation) which do not possess a world-sheet
description in the orientifold language. For the later convenience we give
the massless spectrum of the ${\bf Z}_3$ model in Table I.

{}Let us now go back to the ${\bf Z}_6$ model we have defined in the
beginning of this section. Since the ${\bf Z}_3$ model with the $\Omega$
orientifold projection and the choice of the gauge bundle (\ref{Z_3}) 
is consistent, the ${\bf Z}_6$ model should also be
consistent as long as we make sure that the corresponding ${\bf Z}_2$ model
is consistent. That is, we must choose the action of the ${\bf Z}_2$ twist
on the Chan-Paton factors appropriately. 

{}A consistent solution for the Chan-Paton matrices in the ${\bf Z}_2$
model was given in \cite{PS,GP}. Thus, consider Type IIB on ${\cal
M}_2=T^4/{\bf Z}_2$ (with trivial NS-NS $B$-field), where the generator $R$
of ${\bf Z}_2$ reflects the complex coordinates $z_1,z_2$ on $T^4$:
$Rz_{1,2}=-z_{1,2}$. Next, consider the $\Omega$ orientifold of this
theory. The closed string sector contains (along with the corresponding
${\cal N}=1$ supergravity multiplet) one tensor multiplet and 20
hypermultiplets (4 of which come from the untwisted sector, and the other
16 come from the ${\bf Z}_2$ twisted sector). In the open string sector we have
32 D9-branes plus 32 D5-branes. Let all of the D5-branes sit on top of each
other at the orientifold 5-plane located at the origin of the
orbifold. Then the twisted D9- and D5-brane Chan-Paton matrices 
$\gamma_{9,R}$ and $\gamma_{9,R}$ are given by
\begin{equation}
 \gamma_{9,R}=\gamma_{5,R}={\mbox{diag}}(i{\bf I}_8 , -i{\bf I}_8)~.
\end{equation}   
The massless spectrum of the open string sector consists of gauge vector
supermultiplets of $U(16)_{99}\otimes U(16)_{55}$ as well as the
corresponding charged hypermultiplets transforming in the irreps $2\times 
({\bf 120};{\bf 1})$ (99 sector), $2\times ({\bf 1};{\bf 120})$ (55 sector),
$({\bf 16};{\bf 16})$ (59 sector). This spectrum is summarized in Table I,
and it can be seen to satisfy the
anomaly cancellation condition (\ref{anom}).

{}The ``untwisted'' 99, 55 and 59 sector
matter content in the full ${\bf Z}_6$ model is the same as in the ${\bf
Z}_6$ model of \cite{GJ}. In particular, the orbifold action on the 
Chan-Paton charges is given by the
corresponding twisted Chan-Paton matrices (which are identical for both D9-
and D5-branes in the present setup): 
\begin{eqnarray}\label{CP1}
 &&\gamma_{9,\theta}=\gamma_{5,\theta}=
 {\mbox{diag}}(\alpha {\bf I}_4 , \alpha^{-1} {\bf I}_4, {\bf I}_8)~,\\
 \label{CP2}
 &&\gamma_{9,R}=\gamma_{5,R}={\mbox{diag}}(i{\bf I}_2, -i{\bf I}_2 , 
 i{\bf I}_2, -i{\bf I}_2, i{\bf I}_4, -i{\bf I}_4)~.
\end{eqnarray} 
Here $\theta=g^2$ and $R=g^3$ ($g$ is the generator of the ${\bf Z}_6$
orbifold group). Thus, the gauge group is $[U(4)\otimes U(4)\otimes
U(8)]_{99}$ in the 99 sector. Since we have placed all the D5-branes on
top of each other at the orientifold 5-plane located at the origin, the 55
sector gauge group $[U(4)\otimes U(4)\otimes
U(8)]_{55}$ is the same as in the 99 sector. The matter content of the
``untwisted'' 99, 55 and 59 sectors (these are the sectors that are
perturbative from the orientifold viewpoint) is given in Table I.  

{}We are now ready to determine the ``twisted'' 99, 55 and 59 sector
matter content in the full ${\bf Z}_6$ model. The twisted 99 sector states
are straightforward to determine. We have already discussed the matter
arising in the twisted 99 sector in the ${\bf Z}_3$ model. To obtain the
corresponding states in the ${\bf Z}_6$ model we must project onto the
${\bf Z}_2$ invariant states. Note that the original 9 fixed points can be
organized into 5 linear combinations invariant under the ${\bf Z}_2$ twist
plus 4 linear combinations which pick up a minus sign under the ${\bf Z}_2$
twist. Thus, after taking into account the ${\bf Z}_2$ action on both the
fixed points and Chan-Paton charges, in the twisted 99 sector we have 5
copies of hypermultiplets in the irreps $({\bf 6},{\bf 1},{\bf 1})_{99}$
and $({\bf 1},{\bf 6},{\bf 1})_{99}$, and also 4 copies of hypermultiplets
in the irrep $({\bf 4},{\bf 4},{\bf 1})_{99}$. 

{}The twisted 55 sector is a bit more subtle. Since we have placed all of
the D5-branes at the origin (which is a fixed point of both the ${\bf Z}_3$
and ${\bf Z}_2$ twists), it follows that we have only one copy of
hypermultiplets in irreps $({\bf 6},{\bf 1},{\bf 1})_{55}$
and $({\bf 1},{\bf 6},{\bf 1})_{55}$. (No hypermultiplets in the irrep 
$({\bf 4},{\bf 4},{\bf 1})_{55}$ appear as the origin is invariant under
the ${\bf Z}_2$ twist.) In particular, the other 8 of the original 9 fixed
points of the ${\bf Z}_3$ twist play no role in this discussion as the
twisted 55 states arise due to the {\em local} geometry near the origin.   

{}Finally, the twisted 59 sector contains one copy of hypermultiplets in
the irreps  $({\bf 4},{\bf 1},{\bf 1};{\bf 4},{\bf 1},{\bf 1})_{59}$ and  
$({\bf 1},{\bf 4},{\bf 1};{\bf 1},{\bf 4},{\bf 1})_{59}$. The multiplicity
of these states follows from considerations similar to those in the 55
sector. As to the gauge charges, they are dictated by the fact the twisted
states of this type (due to their origin discussed in the previous section)
can only be charged under the part of the original gauge group (in this
case $SO(32)$) broken by the corresponding (that is, ${\bf Z}_3$) twist.

{}The full massless spectrum of the ${\bf Z}_6$ model is summarized in Table
I (there we also give the massless spectra of the ${\bf Z}_2$ and ${\bf
Z}_3$ models). It is straightforward to check that the anomaly cancellation
condition (\ref{anom}) is satisfied in this model. Note that the twisted
99 and 55
sectors no longer exhibit the naive ``T-duality''. This is due to the fact
that these sectors do {\em not} arise via a straightforward orbifold
reduction of the corresponding ($SO(32)$) gauge theory (with ${\cal N}=2$
supersymmetry in six dimensions). 

{}Thus, the ${\bf Z}_2$ orbifold model of \cite{PS,GP} is perturbative from
the orientifold viewpoint, but is non-perturbative in the heterotic
language. In contrast, the ${\bf Z}_3$ model in Table I is perturbative
from the heterotic viewpoint, but has no perturbative orientifold
description. The ${\bf Z}_6$ model is non-perturbative in both orientifold
and heterotic pictures. Here we should point out that the ${\bf Z}_4$ case
with the $\Omega$ projection (instead of the $\Omega J^\prime$ projection
as in \cite{GJ}) does not appear to be tractable in the above approach as
the action of the ${\bf Z}_4$ twist on the Chan-Paton charges in this case does
not correspond to a perturbative gauge bundle on the heterotic side. Thus,
to understand the ${\bf Z}_4$ case one would need a different approach.

{}We would like to end this section with the following comment. Following
\cite{bij} we could attempt to construct similar non-perturbative
orientifolds with non-trivial NS-NS $B$-field turned on. Unfortunately,
however, we would then lose control over the non-perturbative
(``twisted'') states. Thus, for instance, the ${\bf Z}_3$ model with a 
non-zero NS-NS $B$-field of rank 2 does not appear to have a perturbative
heterotic dual.  
Nonetheless, in the next section we will see that in four
dimensions one can find cases with non-zero NS-NS $B$-field which can be
understood within the present approach.

\section{Four Dimensional Examples}

{}In this section we extend the discussion in the previous section and
obtain non-perturbative orientifolds with ${\cal N}=1$ supersymmetry in
four dimensions. 

{}Thus, consider Type IIB on ${\cal M}_3=T^6/{\bf Z}_6^\prime$ (with
trivial NS-NS $B$-field), where the generators $\theta$ and $R$ of 
the ${\bf Z}_3$ and ${\bf
Z}_2$ subgroups of ${\bf Z}_6^\prime$ act on the complex coordinates on
$T^6$ as follows:
\begin{eqnarray}
 &&\theta z_1 = z_1~,~~~\theta z_2 =\alpha z_2~,~~~\theta z_3=\alpha^{-1}
 z_3 ~,\\
 &&Rz_1 = -z_1~,~~~Rz_2 =-z_2~,~~~Rz_3=z_3 ~,  
\end{eqnarray}
where $\alpha\equiv\exp(2\pi i/3)$. This theory has ${\cal N}=2$
supersymmetry in four dimensions. 

{}Next, consider the $\Omega$ orientifold of this theory. Here $\Omega$
acts as in the smooth Calabi-Yau case. This, in particular, implies that we
must first appropriately blow up (some of) the orbifold singularities
before orientifolding.

{}After orientifolding the closed string sector contains the ${\cal N}=1$
supergravity multiplet, the dilaton plus axion supermultiplet, and also 46
chiral supermultiplets. In particular, 4 of these arise in the untwisted
sector, 18 come from the ${\bf Z}_3$ twisted sector, and ${\bf Z}_2$ and
${\bf Z}_6$ sectors each contribute 12 chiral supermultiplets.

{}As to the open string sector, there are 32 D9-branes and 32
D5-branes. This is required by the untwisted tadpole cancellation
conditions. In the following we will focus on the case where all 32
D5-branes (which wrap the $z_3$ complex direction) are sitting on top of
each other at the orientifold 5-plane located at $z_1=z_2=0$.   

{}The ``untwisted'' 99, 55 and 59 sectors in this orientifold (that is, the
sectors with well defined perturbative description) have been obtained in
\cite{Zw,ibanez,KST}. However, in \cite{KST} it was pointed out that there
are also ``twisted'' 99, 55 and 59 sector contributions one must take into
account. This should be clear from the discussion in the previous
section. In fact, following the approach we have used for constructing the
six dimensional ${\bf Z}_6$ model in the previous section, we can also
obtain the massless states in the twisted open string sectors in the four
dimensional ${\bf Z}_6^\prime$ model as
well. Since the discussion in the four dimensional case parallels quite
closely that for the six dimensional example, we simply summarize the
massless spectrum of the ${\bf Z}_6^\prime$ model in Table II. Let us,
however, mention that the multiplicity 3 in the twisted 55 and 59 sectors 
is due to the fact that there are 3 fixed points in the $z_3$ direction 
which the D5-branes wrap. On the other hand, the original 9 fixed points
relevant in the discussion for the twisted 99 sector decompose into 6
linear combinations invariant under the ${\bf Z}_2$ twist plus 3 linear
combinations that pick up a minus sign under the ${\bf Z}_2$ twist. As to
the orbifold action on the Chan-Paton factors, in this case it is given by
the same Chan-Paton matrices (\ref{CP1}) and (\ref{CP2}) 
as in the six dimensional case.  

{}Note that the four dimensional ${\bf Z}_6^\prime$ 
orbifold model (with trivial
NS-NS $B$-field) given in Table II is chiral. In fact, it is an example of
a chiral ${\cal N}=1$ supersymmetric string vacuum which is
non-perturbative from both the orientifold and heterotic viewpoints. The
spectrum of this model can be seen to be free of non-Abelian gauge
anomalies, and their cancellation in this case in non-trivial. In
particular, the ``untwisted'' open string sectors alone produce an anomaly
free spectrum. The ``twisted'' sector contributions are also anomaly
free. This is why the ``naive'' (that is, perturbative) tadpole cancellation 
conditions for this orientifold give rise to an anomaly free
model. Nonetheless, this does not imply that the spectrum obtained via the
orientifold techniques is complete. In fact, the non-perturbative (that is,
``twisted'') sector contributions are non-trivial, and must be taken into
account.  

{}Before we end this section, we would like to discuss a similar model with
a non-trivial NS-NS $B$-field turned on. Thus, consider the above ${\bf
Z}_6^\prime$ 
orbifold model in the presence of the non-zero NS-NS $B$-field in the
$z_1$ direction\footnote{As we have already mentioned in the previous section,
turning on the $B$-field in the $z_2$ and/or $z_3$ directions would result
in a loss of control over the twisted open string states as they would have
no perturbative description on the heterotic side.}. Thus, let
$T^6=T^2\otimes T^4$, where the two-torus $T^2$ is parametrized by the
complex coordinate $z_1$ (on which the ${\bf Z}_3$ twist $\theta$ acts
trivially), whereas the four-torus $T^4$ is parametrized by the other two 
complex coordinates $z_2,z_3$. Turning on the $B$-field (or rank 2)
corresponding to $T^2$ results in rank reduction in both 99 and 55 sectors
by a factor of 2 \cite{Bij,bij}. On the heterotic side this corresponds to
turning on two Wilson lines on the two non-contractable cycles of $T^2$
such that they reduce the original $SO(32)$ gauge symmetry down to
$SO(16)$ \cite{ZK,BN}. In particular, the first Wilson line corresponding
to, say, the $a$-cycle of $T^2$ breaks $SO(32)$ down to $SO(16)\otimes
SO(16)$. Then the second Wilson line corresponding to the $b$-cycle of
$T^2$ breaks $SO(16)\otimes SO(16)$ to its diagonal $SO(16)_{diag}$
subgroup. This can be viewed as a freely acting ${\bf Z}_2$ orbifold which
permutes the two $SO(16)$ subgroups. In the $SO(16)_{diag}$ language the
${\bf Z}_3$ orbifold action on the ${\mbox{Spin}}(32)/{\bf Z}_2$ degrees of
freedom can then be described in terms of the $8\times 8$ matrices $\{{\bf
W}_1,{\bf W}_\theta,{\bf W}_{\theta^{-1}}\}$, where ${\bf
W}_1={\bf I}_8$, ${\bf W}_\theta={\mbox{diag}}(\alpha{\bf I}_2,
\alpha^{-1}{\bf I}_2,{\bf I}_4)$, and ${\bf W}_{\theta^{-1}}={\bf
W}_\theta^{-1}$. On the orientifold side we then identify the corresponding
$8\times 8$ Chan-Paton matrices via $\{\gamma_1,\gamma_\theta,
\gamma_{\theta^{-1}}\}\equiv\{{\bf W}_1,{\bf W}_\theta,{\bf
W}_{\theta^{-1}}\}$. Such an identification is in fact consistent with the
twisted tadpole cancellation conditions which, in particular, imply that we
must have ${\mbox{Tr}}(\gamma_\theta)=+2$ (instead of
${\mbox{Tr}}(\gamma_\theta)=+4$ as in the case without the $B$-field)
\cite{class}. Thus, we conclude that in this case the heterotic gauge
bundle is indeed perturbative, and we can therefore obtain, say, 
the twisted 99 sector states via the corresponding heterotic computation.

{}The spectrum of the above ${\bf Z}_6^\prime$ model with the $B$-field can be
computed in complete parallel with the case without the $B$-field with the
following straightforward modifications. First, the corresponding twisted
Chan-Paton matrices (which we now write in the $8\times 8$ basis) are given by:
\begin{eqnarray}
 &&\gamma_{9,\theta}=\gamma_{5,\theta}=
 {\mbox{diag}}(\alpha {\bf I}_2 , \alpha^{-1} {\bf I}_2, {\bf I}_4)~,\\
 &&\gamma_{9,R}=\gamma_{5,R}={\mbox{diag}}(i, -i, 
 i, -i, i{\bf I}_2, -i{\bf I}_2)~.
\end{eqnarray}
The gauge group is $[U(2) \otimes U(2) \otimes U(4)]_{99}\otimes 
[U(2) \otimes U(2) \otimes U(4)]_{55}$ (all the D5-branes are sitting on
top of each other at the orientifold 5-plane located at $z_1=z_2=0$). The
spectrum of the open string sector can then be read off the spectrum in
Table II (corresponding to the case without the $B$-field) with the obvious
substitutions for the irreps of the corresponding gauge subgroups. Note,
however, that the multiplicities of (both untwisted as well as twisted) 
59 sector states is now 2 times that shown in Table II. This doubling of
states is due to the rank-2 $B$-field, and was explained in detail in
\cite{bij}. 

{}Before we end this section, let us point out that the approach we have
described in the previous sections can be straightforwardly applied to
study other four dimensional orientifolds with twisted (that is,
non-perturbative) sectors. More concretely, these are the ${\bf Z}_2\otimes
{\bf Z}_6$, ${\bf Z}_3\otimes {\bf Z}_6$ and ${\bf Z}_6\otimes
{\bf Z}_6$ cases whose perturbative (from the orientifold viewpoint)
massless spectra were discussed in \cite{Zw}. (The ${\bf Z}_3\otimes 
{\bf Z}_6$ case was also discussed in \cite{ibanez}.) We will not consider
these models in detail in this paper but will come back to them in
\cite{new} where we intend to discuss a more general class of Type I
compactifications, namely, on Voisin-Borcea orbifolds. Note, however, that
the non-perturbative sectors in the ${\bf Z}_{12}$ model of \cite{ibanez}
are not tractable within the above approach as ${\bf Z}_{12}$ 
contains a ${\bf Z}_4$
subgroup. As we have already mentioned in the previous section, this case
requires techniques other than those employed in this paper.

\acknowledgments
{}I would like to thank Mirjam Cveti{\v c} for valuable discussions. 
This work was supported in part by the grant
NSF PHY-96-02074, 
and the DOE 1994 OJI award. I would also like to thank Albert and 
Ribena Yu for financial support.

\newpage
\begin{figure}[t]
\epsfxsize=16 cm
\epsfbox{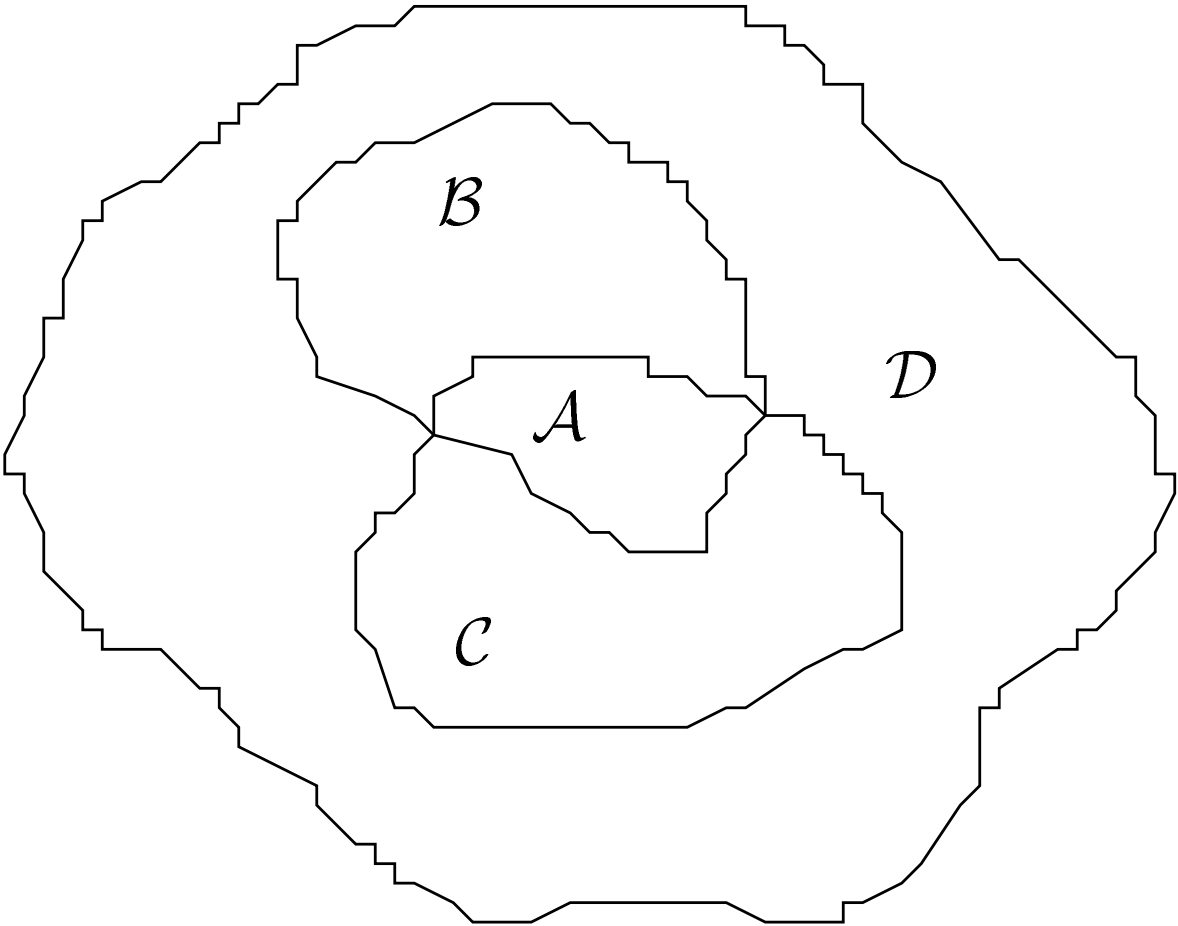}
\bigskip
\caption{A schematic picture of the space of four dimensional ${\cal N}=1$ Type I and
heterotic vacua. The region ${\cal A}\cup{\cal B}$ corresponds to perturbative Type I vacua.
The region ${\cal A}\cup{\cal C}$ corresponds to perturbative heterotic vacua. The vacua
in the region ${\cal A}$ are perturbative from both the Type I and heterotic viewpoints. The region ${\cal D}$ contains both non-perturbative Type I and heterotic vacua.}
\end{figure}

\begin{table}[t]
\begin{tabular}{|c|c|l|c|c|}
Model & Gauge Group & \phantom{Hy} Charged  & Neutral 
& Extra Tensor  \\
     &             &Hypermultiplets & Hypermultiplets
&Multiplets \\
\hline
${\bf Z}_2$ & $U(16)_{99} \otimes U(16)_{55}$ & 
 $2 ({\bf 120};{\bf 1})_U$ & $20$
& $0$ \\
               &                       & $2 ({\bf 1};{\bf 120})_U$ & & \\
               &                       & $({\bf 16};{\bf 16})_U$ & & \\
\hline
${\bf Z}_3$ & $[U(8) \otimes SO(16)]_{99}$ & $({\bf 28},{\bf 1})_U$ & $20$
& $0$ \\
& & $({\bf 8},{\bf 16})_U$ & & \\
& & $9\times ({\bf 28},{\bf 1})_T$ & & \\
\hline
${\bf Z}_6$ & $[U(4) \otimes U(4) \otimes U(8)]_{99}\otimes$ 
& $({\bf 6},{\bf 1},{\bf 1};{\bf 1},{\bf 1},{\bf 1})_U$ & $20$ & $0$ \\
& $[U(4) \otimes U(4) \otimes U(8)]_{55}$&
$({\bf 1},{\bf 6},{\bf 1};{\bf 1},{\bf 1},{\bf 1})_U$  & & \\
& & $({\bf 4},{\bf 1},{\bf 8};{\bf 1},{\bf 1},{\bf 1})_U$ & & \\
& & $({\bf 1},{\bf 4},{\bf 8};{\bf 1},{\bf 1},{\bf 1})_U$ & & \\
& & $({\bf 1},{\bf 1},{\bf 1};{\bf 6},{\bf 1},{\bf 1})_U$ & & \\
& & $({\bf 1},{\bf 1},{\bf 1};{\bf 1},{\bf 6},{\bf 1})_U$ & & \\
& & $({\bf 1},{\bf 1},{\bf 1};{\bf 4},{\bf 1},{\bf 8})_U$ & & \\
& & $({\bf 1},{\bf 1},{\bf 1};{\bf 1},{\bf 4},{\bf 8})_U$ & & \\
& & $({\bf 4},{\bf 1},{\bf 1};{\bf 4},{\bf 1},{\bf 1})_U$ & & \\
& & $({\bf 1},{\bf 4},{\bf 1};{\bf 1},{\bf 4},{\bf 1})_U$ & & \\
& & $({\bf 1},{\bf 1},{\bf 8};{\bf 1},{\bf 1},{\bf 8})_U$ & & \\
& & $5\times ({\bf 6},{\bf 1},{\bf 1};{\bf 1},{\bf 1},{\bf 1})_T$ & & \\
& & $5\times ({\bf 1},{\bf 6},{\bf 1};{\bf 1},{\bf 1},{\bf 1})_T$ & & \\
& & $4\times ({\bf 4},{\bf 4},{\bf 1};{\bf 1},{\bf 1},{\bf 1})_T$ & & \\
& & $({\bf 1},{\bf 1},{\bf 1};{\bf 6},{\bf 1},{\bf 1})_T$ & & \\
& & $({\bf 1},{\bf 1},{\bf 1};{\bf 1},{\bf 6},{\bf 1})_T$ & & \\
& & $({\bf 4},{\bf 1},{\bf 1};{\bf 4},{\bf 1},{\bf 1})_T$ & & \\
& & $({\bf 1},{\bf 4},{\bf 1};{\bf 1},{\bf 4},{\bf 1})_T$ & & \\
\hline
\end{tabular}
\caption{The massless spectra of the six dimensional Type IIB orientifolds
on $T^4/{\bf Z}_N$ for $N=2,3,6$.
The semi-colon in the column ``Charged Hypermultiplets'' separates $99$ and 
$55$ representations. The subscript ``$U$'' indicates that the
corresponding (``untwisted'') state is perturbative from the orientifold
viewpoint. The subscript ``$T$'' indicates that the
corresponding (``twisted'') state is non-perturbative from the orientifold
viewpoint. The $U(1)$ charges are not shown, and by ``neutral''
hypermultiplets we mean that the corresponding states are not charged
under the non-Abelian subgroups.}
\end{table}

\begin{table}[t]
\begin{tabular}{|c|c|l|c|}
Model & Gauge Group & \phantom{Hy} Charged  & Neutral 
 \\
     &             &Chiral Multiplets & Chiral Multiplets
\\
\hline
${\bf Z}_6^\prime$ & $[U(4) \otimes U(4) \otimes U(8)]_{99}\otimes$ 
& $({\bf 1},{\bf 1},{\bf 28};{\bf 1},{\bf 1},{\bf 1})_U$ & $U:~~~4$ \\
& $[U(4) \otimes U(4) \otimes U(8)]_{55}$&
$({\bf 1},{\bf 1},{\overline {\bf 28}};
 {\bf 1},{\bf 1},{\bf 1})_U$  & ${\bf Z}_3:~~~18$ \\
& & $({\bf 4},{\bf 4},{\bf 1};{\bf 1},{\bf 1},{\bf 1})_U$ & ${\bf
Z}_2:~~~12$   \\
& & $({\overline {\bf 4}},{\overline {\bf 4}},{\bf 1};
 {\bf 1},{\bf 1},{\bf 1})_U$ & ${\bf
Z}_6:~~~12$  \\
& & $({\bf 4},{\bf 1},{\bf 8};
 {\bf 1},{\bf 1},{\bf 1})_U$ &  \\
& & $({\bf 1},{\overline {\bf 4}},{\overline {\bf 8}};
 {\bf 1},{\bf 1},{\bf 1})_U$ &  \\
& & $({\overline {\bf 6}},{\bf 1},{\bf 1};
 {\bf 1},{\bf 1},{\bf 1})_U$ &  \\
& & $({\bf 1},{\bf 6},{\bf 1};
 {\bf 1},{\bf 1},{\bf 1})_U$ &  \\
& & $({\overline {\bf 4}},{\bf 1},{\bf 8};
 {\bf 1},{\bf 1},{\bf 1})_U$ &  \\
& & $({\bf 1},{\bf 4},{\overline {\bf 8}};
 {\bf 1},{\bf 1},{\bf 1})_U$ &  \\
& & $({\bf 4},{\overline {\bf 4}},{\bf 1};
 {\bf 1},{\bf 1},{\bf 1})_U$ &  \\
& & Same as above with $99\leftrightarrow 55$ & \\
& & $({\bf 4},{\bf 1},{\bf 1};{\bf 4},{\bf 1},{\bf 1})_U$ & \\
& & $({\bf 1},{\bf 4},{\bf 1};{\bf 1},{\bf 1},{\bf 8})_U$ &  \\
& & $({\bf 1},{\bf 1},{\bf 8};{\bf 1},{\bf 4},{\bf 1})_U$ & \\
& & $({\overline {\bf 4}},{\bf 1},{\bf 1};{\bf 1},{\bf 1},
 {\overline {\bf 8}})_U$ & \\
& & $({\bf 1},{\overline {\bf 4}},{\bf 1};{\bf 1},{\overline {\bf 4}},
 {\bf 1})_U$ &  \\
& & $({\bf 1},{\bf 1},{\overline {\bf 8}};{\overline {\bf 4}},{\bf 1},
 {\bf 1})_U$ & \\
& & $6\times ({\bf 4},{\overline {\bf 4}},{\bf 1};
 {\bf 1},{\bf 1},{\bf 1})_T$ &  \\
& & $3\times ({\overline {\bf 4}},{\bf 4},{\bf 1};
 {\bf 1},{\bf 1},{\bf 1})_T$ &  \\
& & $6\times ({\overline {\bf 6}},{\bf 1},{\bf 1};
 {\bf 1},{\bf 1},{\bf 1})_T$ &  \\
& & $3\times ({\bf 6},{\bf 1},{\bf 1};
 {\bf 1},{\bf 1},{\bf 1})_T$ &  \\
& & $6\times ({\bf 1},{\bf 6},{\bf 1};
 {\bf 1},{\bf 1},{\bf 1})_T$ &  \\
& & $3\times ({\bf 1},{\overline {\bf 6}},{\bf 1};
 {\bf 1},{\bf 1},{\bf 1})_T$ &  \\
& & $3\times ({\bf 1},{\bf 1},{\bf 1};
 {\bf 4},{\overline {\bf 4}},{\bf 1})_T$ &  \\
& & $3\times ({\bf 1},{\bf 1},{\bf 1};
 {\overline {\bf 6}},{\bf 1},{\bf 1})_T$ &  \\
& & $3\times ({\bf 1},{\bf 1},{\bf 1};{\bf 1},{\bf 6},{\bf 1})_T$ &  \\

& & $3\times ({\overline {\bf 4}},{\bf 1},{\bf 1};{\overline {\bf 4}},
 {\bf 1},{\bf 1})_T$ & \\
& & $3\times ({\bf 1},{\bf 4},{\bf 1};{\bf 1},{\bf 4},{\bf 1})_T$ &  \\
\hline
\end{tabular}
\caption{The massless spectrum of the four dimensional Type IIB orientifold
on $T^6/{\bf Z}_6^\prime$.
The semi-colon in the column ``Charged Chiral Multiplets'' separates $99$ and 
$55$ representations. The subscript ``$U$'' indicates that the
corresponding (``untwisted'') state is perturbative from the orientifold
viewpoint. The subscript ``$T$'' indicates that the
corresponding (``twisted'') state is non-perturbative from the orientifold
viewpoint. The $U(1)$ charges are not shown, and by ``neutral''
chiral multiplets we mean that the corresponding states are not charged
under the non-Abelian subgroups.}
\end{table}


\begin{references}

\bibitem{PS} G. Pradisi and A. Sagnotti, Phys. Lett. {\bf B216} (1989) 59;\\
M. Bianchi and A. Sagnotti, Phys. Lett. {\bf B247} (1990) 517; Nucl. Phys. 
{\bf B361} (1991) 539. 

\bibitem{GP} E.G. Gimon and J. Polchinski, Phys. Rev. {\bf D54} (1996) 1667.

\bibitem{GJ}E.G. Gimon and C.V. Johnson, Nucl. Phys. {\bf B477} (1996) 715.\\
A. Dabholkar and J. Park, Nucl. Phys. {\bf B477} (1996) 701.

\bibitem{bij} Z. Kakushadze, G. Shiu and S.-H.H. Tye, Phys. Rev. {\bf D58}
(1998) 086001.

\bibitem{BL} M. Berkooz and R.G. Leigh, Nucl. Phys. {\bf B483} (1997) 187.

\bibitem{Sagnotti} C. Angelantonj, M. Bianchi, G. Pradisi, A. Sagnotti and 
Ya.S. Stanev, Phys. Lett. {\bf B385} (1996) 96.

\bibitem{KS} Z. Kakushadze and G. Shiu, Phys. Rev. {\bf D56} (1997) 3686; 
Nucl. Phys. {\bf B520} (1998) 75.

\bibitem{Zw} G. Zwart, Nucl. Phys. {\bf B526} (1998) 378.

\bibitem{ibanez} G. Aldazabal, A. Font, L.E. Ib{\'a}{\~n}ez and G. Violero,
Nucl. Phys. {\bf B536} (1998) 29.

\bibitem{KST} Z. Kakushadze, G. Shiu and S.-H.H. Tye, 
Nucl. Phys. {\bf B533} (1998) 25.

\bibitem{ZK} Z. Kakushadze, Nucl. Phys. {\bf B512} (1998) 221.

\bibitem{223} Z. Kakushadze, Phys. Lett. {\bf B434} (1998) 269.

\bibitem{class} Z. Kakushadze, Nucl. Phys. {\bf B535} (1998) 311.

\bibitem{pol} J. Polchinski, Phys. Rev. {\bf D55} (1997) 6423. 

\bibitem{blum} J.D. Blum, Nucl. Phys. {\bf B486} (1997) 34.

\bibitem{anomalies}
M.B. Green, J.H. Schwarz and P.C. West, Nucl. Phys. {\bf B254} (1985) 327;\\
J. Erler, J. Math. Phys. {\bf 35} (1994) 1819.

\bibitem{Bij} 
M. Bianchi, G. Pradisi and A. Sagnotti, Nucl. Phys. {\bf B376} (1992) 365.

\bibitem{BN} Z. Kakushadze, Nucl. Phys. {\bf B544} (1999) 265.

\bibitem{new} Z. Kakushadze, to appear. 



\end{references}
\end{document}